\documentstyle[12pt]{article}

\author{V. A. Tsokur, Yu. M. Zinoviev
        \thanks{E-mail address: ZINOVIEV@MX.IHEP.SU} \\
        {\it Institute for High Energy Physics} \\
        {\it Protvino, Moscow Region, 142284, Russia}}
\title{Dual versions of extended supergravities}
\date{November 1994}

\begin{document}
\maketitle

\begin{abstract}
   Recently, using the model of $N=2$ supergravity --- vector
multiplets interaction with the scalar field geometry
$SU(1,m)/SU(m)\otimes U(1)$ as an example, we have shown that even when
the scalar field geometry is fixed, one can have a whole family of the
Lagrangians, which differ by vector field duality transformation. In
this paper we carry out the construction of such families for the
case of $N=3$ and $N=4$ supergravities, the scalar field geometry
being $SU(3,m)/SU(3)\otimes SU(m)\otimes U(1)$ and $SU(1,1)/U(1)\otimes
O(6,m)/O(6)\otimes O(m)$, correspondingly. Moreover, it turns out that
these families contain, as a partial case, the models describing the
interaction of arbitrary number of vector multiplets with our hidden
sectors, admitting spontaneous supersymmetry breaking without
a cosmological term.
\end{abstract}

\newpage

\section{Introduction}

  As is well known (see e.g. \cite{Der84}), scalar fields in extended
supergravities usually describe some non-linear $\sigma $-models of the
form $G/H$, where $G$ is some noncompact group, and $H$ --- its
maximal compact subgroup. In this, the group $G$ would be the global
symmetry of the whole Lagrangian, but in many models it turns out
that the realization of this group on the vector fields includes
duality transformations. This leads to the fact that only a part of
the group $G$ appears to be the symmetry of the Lagrangian, the rest
being the symmetry of the equations of motion only. In turn, the
invariance of the equations of motion leads to the existence of the
whole family of the Lagrangians (we will call them dual versions)
with the same scalar field geometry. In our recent paper
\cite{Zin94}, we have demonstrated this fact on the simplest example
--- $N=2$ supergravity with vector multiplets and scalar field
geometry $SU(1,m)/SU(m)\otimes U(1)$. Moreover, a deep connection between
different dual versions and the problem of spontaneous supersymmetry
breaking appeared.

   In this paper we consider the generalization of such construction
to the case of $N=3$ and $N=4$ supergravities. For completeness, in
the next section we reproduce our $N=2$ example, mentioned above. Then
we show that it is easy to generalize this result to the case of $N=3$
supergravity with arbitrary number of vector multiplets and scalar
field geometry $SU(3,m)/SU(3)\otimes SU(m)\otimes U(1)$. Moreover, it
turns out that the Lagrangian constructed contains, as a partial case,
model \cite{Tso94}, describing an interaction of "matter" vector
multiplets with the hidden sector \cite{Zin86b}, admitting spontaneous
supersymmetry breaking without a cosmological term.

   The $N=4$ case appears to be more difficult. The reason is that the
$N=4$ supergravity itself \cite{Das77,Cre77,Cre78} already contains
two scalar fields, corresponding to the non-linear $\sigma $-model
$SU(1,1)/U(1)$. So, we started with the pure $N=4$ supergravity and
found the most general Lagrangian for this theory. Then we have
managed to construct the family of the Lagrangians, which gives the
most general (to our knowledge) interaction of $N=4$ supergravity
with vector multiplets, scalar field geometry for the whole family
being $SU(1,1)/U(1)\otimes O(6,m)/O(6)\otimes O(m)$. As in the $N=3$ case,
it turns out that as a partial case this family contains   model
\cite{Tso94a}, describing an interaction of vector multiplets with
hidden sector \cite{Zin87}.

\section{$N=2$ supergravity}

   In this section we give a brief description of our model for
general interaction of $N=2$ supergravity with vector multiplets, the
scalar field geometry being $SU(1,m)/SU(m)\otimes U(1)$. The simplest
way to describe this model is to introduce, apart from the graviton
$e_{\mu r}$ and gravitini $\Psi _{\mu i}$, $i=1,2$, $m+1$ vector
multiplets $\{A_\mu {}^a, \Omega _{ia}, z^a\}$, $a=0,1,...m$. The
following constraints correspond to the model with required geometry:
\begin{equation}
 \bar{z}_a \cdot  z^a = - 2,\qquad z^a \cdot  \Omega _{ia} = 0. \label{c1}
\end{equation}
In this, the theory has local axial $U(1)$ invariance with composite
gauge field $U_\mu  = (\bar{z}_a \partial _\mu  z^a)$. The corresponding
covariant derivatives look like, e.g.,
\begin{eqnarray}
 {\cal D}_\mu  z^a &=& \partial _\mu  z^a + \frac{1}{2} (\bar{z} \partial
_\mu  z) z^a, \qquad \bar{z}_a {\cal D}_\mu  z^a = 0, \nonumber \\
 {\cal D}_\mu  \eta _i &=& D_\mu  \eta _i - \frac{1}{4} (\bar{z} \partial
_\mu  z) \eta _i.
\end{eqnarray}

   The crucial point is that this local $U(1)$ invariance and
constraints (\ref{c1}) do not determine the form of the Lagrangian and
supertransformations unambiguously. Indeed, one can make the following
general ansatz for the supertransformations:
\begin{eqnarray}
 \delta  e_{\mu r} &=& i (\bar{\Psi }_\mu {}^i \gamma _r \eta _i),
\nonumber \\
 \delta  \Psi _{\mu i} &=& 2 {\cal D}_\mu  \eta _i - \frac{i}{2}
\varepsilon _{ij} \frac{z^a E_{a\alpha }
(\sigma A)^\alpha }{(z g z)} \gamma _\mu  \eta ^j, \nonumber \\
 \delta  A_m{}^\alpha  &=& \varepsilon ^{ij} (\bar{\Psi }_{\mu i} z^a
K_a{}^\alpha  \eta _j) + i (\bar{\Omega }^{ia}
\gamma _\mu  K_a{}^\alpha  \eta _i) , \\
 \delta  \Omega _{ia} &=& - \frac{1}{2} \left[ E_{a\alpha } (\sigma
A)^\alpha  - \frac{g_{ab} z^b
(z^c E_{c\alpha } (\sigma A)^\alpha )}{(z g z)} \right] \eta _i - i
\varepsilon _{ij} \hat{\cal D} z^a
\eta ^j, \nonumber \\
 \delta  x^a &=& \varepsilon ^{ij} (\bar{\Omega }_{ia} \eta _j) \qquad
\delta  y^a = \varepsilon ^{ij}
(\bar{\Omega }_{ia} \gamma _5 \eta _j), \nonumber
\end{eqnarray}
where $E_{a\alpha }$, $K_a{}^\alpha $, $\alpha =0,1,...m$ and $g_{ab}$ are
constant matrices and then find the corresponding Lagrangian:
\begin{eqnarray}
 L &=& - \frac{1}{2} R + \frac{i}{2} \varepsilon ^{\mu \nu \rho \sigma }
\bar{\Psi }_\mu {}^i \gamma _5 \gamma _\nu
{\cal D}_\rho  \Psi _{\sigma i} + \frac{i}{2} \bar{\Omega }^{ia} \hat{\cal
D} \Omega _{ia} +
\frac{1}{2} {\cal D}_\mu  \bar{z}_a {\cal D}_\mu  z^a - \nonumber \\
  && - \frac{1}{4} A_{\mu \nu }{}^\alpha  E_{a\alpha } \bar{E}^a{}_\beta
A_{\mu \nu }{}^\beta  +
\frac{1}{4} \left[ \frac{(z E A_{\mu \nu })(z E (A_{\mu \nu } + \gamma _5
\tilde{A}_{\mu \nu }))}{(z g z)} + h.c. \right] - \nonumber \\
  && - \frac{1}{2} \varepsilon ^{ij} \bar{\Psi }_{\mu i} \frac{\bar{z}_a
\bar{E}^a{}_\alpha
(A^{\mu \nu } - \gamma _5 \tilde{A}^{\mu \nu })^\alpha }{(\bar{z} \bar{g}
\bar{z})} \Psi _{\nu j} -
\frac{1}{2} \varepsilon ^{ij} \bar{\Omega }^{ia} \gamma ^\mu  \gamma ^\nu
{\cal D}_\nu  z^a \Psi _{\mu j} + \nonumber \\
  && + \frac{i}{4} \bar{\Omega }^{ia} \gamma ^\mu  \left[ E_{a\alpha }
(\sigma A)^\alpha  -
\frac{g_{ab}z^b (z E (\sigma A))}{(z g z)} \right] \Psi _{\mu i}
\end{eqnarray}
In this, the requirements of the closure of the superalgebra and the
invariance of the Lagrangian give:
\begin{equation}
 \bar{K}^{a\alpha } E_{a\beta } = \delta ^\alpha {}_\beta ,  \qquad g_{ab}
= K_a{}^\alpha  E_{b\alpha }, \qquad
E_{a[\alpha } \bar{E}^a{}_{\beta ]} = 0.
\end{equation}
The first two equations allow one to express $K_a{}^\alpha $ and $g_{ab}$
in terms of the $E_{a\alpha }$ while the last one turns out to be the only
constraint on $E_{a\alpha }$. Note that in such model vector fields
$A_\mu {}^\alpha $ and spinor $\Omega _{ia}$ and scalar $z^a$ fields carry
different kind of indices exactly as in the general construction of
\cite{D'A91}.

   Thus, we  have really, obtained a whole family of the Lagrangians with
the same scalar field geometry. In \cite{Zin94} we have shown that as
a partial case such family contains a model corresponding to the
interaction of arbitrary number of vector multiplets with  hidden
sector \cite{Zin86}, admitting spontaneous supersymmetry breaking
without a cosmological term. So, it is the choice of the dual version
that determines the possibility of spontaneous supersymmetry breaking,
while the scalar field geometry determines the pattern of such
breaking, i.e. the structure of soft breaking terms that are generated
after symmetry breaking had taken place. It is this close connection of
the existence of dual versions and the problem of spontaneous
supersymmetry breaking that makes an investigation of dual version for
other extended supergravities interesting.

\section{$N=3$ supergravity}

   The $N=3$ vector multiplets contain complex scalar fields $z_i$,
$i=1,2,3$, which are transformed under the triplet representation of
$SU(3)$ group. So the only natural candidate for the scalar field
geometry, corresponding to the $N=3$ supergravity --- vector multiplet
interaction is the non-linear $\sigma $-model $SU(3,m)/SU(3)\otimes
SU(m)\otimes U(1)$.
Such a model was indeed  constructed \cite{Cas86,Zin86b} some time
ago. It turned out that due to reality of the vector fields only the
real subgroup $O(3,m)$ appeared to be the global symmetry of the
Lagrangian, while the rest of the $SU(3,m)$ group (containing vector
field duality transformations) was the symmetry of the equations of
motion only. But this means that there exists a whole family of the
Lagrangians with the same scalar field geometry but with different
vector fields coupling. It is this general coupling that we are going to
construct in this section.

   Let us introduce the following set of fields: graviton $e_{\mu r}$,
gravitini $\Psi _{\mu i}$, Majorana spinor $\rho $ and $m+3$ vector
multiplets $\{A_\mu {}^a, \Omega _{ia}, \lambda ^a, z_i{}^a \}$,
where $a=1,2,...m+3$ with the
signature $(---,++...+)$. Then in order to have the model with the
required scalar field geometry one has to impose the following
constraints on the scalar and spinor fields:
\begin{equation}
 z_i{}^a \bar{z}_a{}^j = - 2 \delta _i{}^j, \qquad z_i{}^a \Omega _{ja} =
\bar{z}^i{}_a \lambda ^a = 0. \label{c2}
\end{equation}
In this, the theory has local $SU(3)\otimes U(1)$ invariance with
composite gauge fields, e.g.
\begin{equation}
 {\cal D}_\mu  z_i{}^a = \partial _\mu  z_i{}^a - \frac{1}{2} (z_i
\partial _\mu  \bar{z}^j)
z_j{}^a, \qquad \bar{z}^i {\cal D}_\mu  z_j = 0.
\end{equation}

   Now, by the analogy with the $N=2$ case, we will try the following
ansatz for the supertransformations, compatible with  constraints
(\ref{c2}) and $SU(3)\otimes U(1)$ invariance:
\begin{eqnarray}
 \delta e_{\mu r} &=& i (\bar{\Psi }_\mu {}^i \gamma _r \eta _i),
\nonumber \\
 \delta \Psi _{\mu i} &=& 2 {\cal D}_\mu  \eta _i + \frac{i}{2}
\varepsilon _{ijk} (g^{-1})^{jl}
[z_l{}^a M_a{}^\alpha  (\sigma A)_\alpha  ] \gamma _\mu  \eta ^k,
\nonumber \\
 \delta A_\mu {}^\alpha  &=& - \varepsilon ^{ijk} (\bar{\Psi }_{\mu i}
z_j{}^a K_a{}^\alpha  \eta _k) +
\frac{i}{\sqrt{2}} \bar{\rho } z_i{}^a K_a{}^\alpha  \gamma _\mu  \eta _i
+ i (\bar{\Omega }_{ia}
\bar{K}^{a\alpha } \gamma _\mu  \eta _i),  \\
 \delta \Omega _{ia} &=& - \frac{1}{2} \{ M_a{}^\alpha  (\sigma A)_\alpha
- g_{ab} z_j{}^b
(g^{-1})^{jk} [z_k{}^c M_c{}^\alpha  (\sigma A)_\alpha  ] \} \eta _i + i
\varepsilon _{ijk} \gamma ^\mu
{\cal D}_\mu  z_j{}^a \eta _k \nonumber, \\
 \delta \rho  &=& - \frac{1}{\sqrt{2}} (g^{-1})^{ij} [z_j{}^a M_a{}^\alpha
 (\sigma A)_\alpha  ]
\eta _i, \qquad \delta \lambda ^a = - i \gamma ^\mu  {\cal D}_\mu
\bar{z}^i{}_a \eta _i, \nonumber \\
 \delta \varphi _i{}^a &=& (\bar{\lambda }^a \eta _i) + \varepsilon _{ijk}
(\bar{\Omega }^{ja} \eta ^k), \qquad
\delta \pi _i{}^a = -(\bar{\lambda }^a \gamma _5 \eta _i) + \varepsilon
_{ijk} (\bar{\Omega }^{ja} \gamma _5 \eta ^k),
\nonumber
\end{eqnarray}
where $K_a{}^\alpha $, $M_a{}^\alpha $, $\alpha =1,2,...m+3$ and $g_{ab}$
are constant matrices, while $g_{ij} = z_i{}^a g_{ab} z_j{}^b$. Note,
that a rather complicated structure for $\delta \Omega _{ia}$ was
chosen so that $z_i{}^a \delta \Omega _{ja} = 0$. In this, the
requirement of the closure of the superalgebra on the bosonic fields
leads to:
\begin{equation}
 \bar{K}^a{}_\alpha  M_a{}^\beta  = \delta _\alpha {}^\beta ,  \qquad
K_a{}^\alpha  = g_{ab} \bar{K}^{b\alpha }.
\end{equation}

   By rather long but straightforward calculations one can construct the
corresponding Lagrangian:
\begin{eqnarray}
 L &=& - \frac{1}{2} R + \frac{i}{2} \varepsilon ^{\mu \nu \rho \sigma }
\bar{\Psi }_\mu {}^i \gamma _5 \gamma _\nu
{\cal D}_\rho  \Psi _{\sigma i} + \frac{i}{2} \bar{\rho } \hat{\cal D}
\rho  + \nonumber \\
 && + \frac{i}{2} \bar{\Omega }^i \hat{\cal D} \Omega _i + \frac{1}{2}
\bar{\lambda }
\hat{\cal D} \lambda  + \frac{1}{2} {\cal D}_\mu  \bar{z}^i {\cal D}_\mu
z_i - \frac{1}{4} M_a{}^\alpha  \bar{M}^{a\beta } A_{\mu \nu }{}^\alpha
 A_{\mu \nu }{}^\beta  + \nonumber \\
 && + \frac{1}{4} M_a{}^\alpha  z_j{}^a (g^{-1})^{jk} z_k{}^b M_b{}^\beta
A_{\mu \nu }{}^\alpha  (A_{\mu \nu } + i \tilde{A}_{\mu \nu })^\beta  +
h.c. + \nonumber \\
 && + \frac{1}{2} \varepsilon ^{ijk} \bar{\Psi }_{\mu i}
(\bar{g}^{-1})_{jl} \bar{z}^l{}_a \bar{M}^a{}_\alpha  (A_{\mu \nu }
 - \gamma _5 \tilde{A}_{\mu \nu })^\alpha  \Psi _{\nu k} +
\nonumber \\
 && + \frac{i}{2\sqrt{2}} \bar{\rho } \gamma ^\mu  (g^{-1})^{ij} z_j{}^a
M_a{}^\alpha (\sigma A)_\alpha  \Psi _{\mu i} + \nonumber \\
 && + \frac{i}{4} \bar{\Omega }^{ia} \gamma ^\mu  [ M_a{}^\alpha  (\sigma
A)_\alpha  - g_{ab} z_j{}^b
(g^{-1})^{jk} z_k{}^c M_c{}^\alpha  (\sigma A)_\alpha  ] \Psi _{\mu i} +
\nonumber \\
 && + \frac{1}{2} \varepsilon ^{ijk} \bar{\Omega }_{ia} \gamma ^\mu
\gamma ^\nu  {\cal D}_\nu  z_j{}^a
\Psi _{\mu k} - \frac{1}{2} \bar{\lambda }^a \gamma ^\mu  \gamma ^\nu
{\cal D}_\nu  \bar{z}^i{}_a \Psi _{\mu i} - \nonumber \\
 && - \frac{1}{2\sqrt{2}} \bar{\lambda }^a [ M_a{}^\alpha  (\sigma
A)_\alpha  - g_{ab} z_j{}^b (g^{-1})^{jk} z_k{}^c M_c{}^\alpha
 (\sigma A)_\alpha  ] \rho  - \nonumber \\
 && - \frac{1}{2} \bar{\lambda }^a (g^{-1})^{ij} z_j{}^b M_b{}^\alpha
(\sigma A)_\alpha \Omega _{ia}.
\end{eqnarray}
This Lagrangian will be invariant under the supertransformations
given above provided:
\begin{equation}
 g_{ab} = K_a{}^\alpha  M_b{}^\alpha ,  \qquad M_a{}^{[\alpha }
\bar{M}^{a|\beta ]} = 0.
\end{equation}
These relations together with ones from the requirement of the closure
of the superalgebra allow one to express $K_a{}^\alpha $ and $g_{ab}$ in
terms of $M_a{}^\alpha $, the last relation being the only constraint on
it.

   Thus, we indeed have obtained a whole family of Lagrangians and one
can easily check that previously known one \cite{Zin86b}
corresponds to the trivial case $M_a{}^\alpha  = \delta _a{}^\alpha $. A
less trivial (and physically more interesting) case is  model \cite{Tso94},
describing the interaction of the arbitrary number of vector multiplets
with hidden sector \cite{Zin86b} (recall, that it is just the dual
version of the system $N=3$ supergravity with three vector
multiplets). The corresponding matrix $M_a{}^\alpha $ looks like:
\begin{equation}
 M = \left( \begin{array}{c|c|c} I_{3\times 3} & \gamma _5 \times
I_{3\times 3} & 0_{3\times (m-3)} \\ \hline I_{3\times 3} & -
 \gamma _5 \times  I_{3\times 3} & 0_{3\times (m-3)} \\ \hline
0_{(m-3)\times 3} & 0_{(m-3)\times 3} & I_{(m-3)\times (m-3)} \end{array}
\right).
\end{equation}
The only difficulty arises from the fact that matrix $g_{ab}$ and hence
$g_{ij}$ turns out to be degenerate in this case (this is the reason
for the enhancement of the global symmetry and other peculiar features
of the model). So, one has to make some kind of regularization for the
matrix $M_a{}^\alpha $ to avoid singularities, keeping only the terms
which survive when regularization parameter goes to zero and making field
rescaling if necessary. We have explicitly checked that one indeed can
reproduce all the formulas from \cite{Tso94} in this way.

\section{$N=4$ supergravity}

   As we have already mentioned, $N=4$ case is more complicated due to
the presence of two scalar fields in the supergravity multiplets.
These fields parameterize non-linear $\sigma $-model $SU(1,1)/U(1)$, in
this, the group $SU(1,1)$ fails to be the global symmetry of the whole
Lagrangian including vector field terms. This leads to the existence
of different dual versions for such theory, the best known examples
being so called $O(4)$ \cite{Das77,Cre77} and $SU(4)$ \cite{Cre78}
supergravities.

   To describe the most general form of $N=4$ supergravity let us
introduce the following fields: graviton $e_{\mu r}$, gravitini $\Psi
_{\mu i}$, $i=1,2,3,4$, vector fields $A_\mu {}^a$, $a=1,2,3,4,5,6$,
Majorana spinors $\lambda _\alpha $, $\alpha =0,1$ and a couple of
complex scalars $z^\alpha $. The scalars and spinors satisfy the usual
constraints:
\begin{equation}
 z^\alpha  \bar{z}_\alpha  = - 2, \qquad z^\alpha  \lambda _\alpha  = 0
\end{equation}
and the theory has local $U(1)$ invariance with the composite gauge
field. Now one can choose the following general ansatz for the
supertransformations:
\begin{eqnarray}
 \delta e_{\mu r} &=& i (\bar{\Psi }_\mu  \gamma _r \eta ), \nonumber \\
 \delta \Psi _\mu  &=& 2 {\cal D}_\mu  \eta  - \frac{i}{4} (\sigma A)^m
E_{ma} \tau ^a \eta ,  \nonumber \\
 \delta A_\mu {}^m &=& \bar{\Psi }_\mu  z^\alpha  K_\alpha {}^{ma}
\bar{\tau }^a \eta  - i \bar{\lambda }^\alpha  \gamma _\mu
K_\alpha {}^{ma} \bar{\tau }^a \eta ,  \\
 \delta \lambda _\alpha  &=& - \frac{1}{4} \varepsilon _{\alpha \beta }
z^\beta  (\sigma A)^m E_{ma} \bar{\tau }^a \eta  - i \gamma ^\mu
{\cal D}_\mu  z^\alpha  \eta ,  \nonumber \\
 \delta \bar{z}_\alpha  &=& 2 (\bar{\lambda }_\alpha  \eta ), \nonumber
\end{eqnarray}
where $K_\alpha {}^{ma}$ is a constant matrix, while $E_{ma}$ is a
function of $z_\alpha $ with the axial charge equal to that of
$\bar{z}_\alpha $. Besides, we introduced six antisymmetric
matrices $(\tau ^a)_{[ij]}$, such that
\begin{equation}
 (\bar{\tau }^a)^{ij} = \frac{1}{2} \varepsilon ^{ijkl} (\tau ^a)_{kl},
\qquad (\tau ^a)_{ij}
(\bar{\tau }^b)^{jk} + ( a \leftrightarrow  b) = - 2 \delta _i{}^k.
\end{equation}
The requirement of the closure of the superalgebra on the bosonic
fields leads to:
\begin{equation}
 K_\alpha {}^{ma} = \varepsilon _{\alpha \beta } \bar{K}^{\beta ,ma,}
\qquad K^{ma} E_{na} = z^\alpha
K_\alpha {}^{ma} E_{na} = \delta ^m{}_n.
\end{equation}
The corresponding ansatz for the Lagrangian looks like:
\begin{eqnarray}
 L &=& - \frac{1}{2} R + \frac{i}{2} \varepsilon ^{\mu \nu \rho \sigma }
\bar{\Psi }_\mu  \gamma _5 \gamma _\nu  {\cal
D}_\rho  \Psi _\sigma  + \frac{i}{2} \bar{\lambda }^\alpha  \hat{\cal D}
\lambda _\alpha  + \frac{1}{2} {\cal
D}_\mu  z^\alpha  {\cal D}_\mu  \bar{z_\alpha } + \nonumber \\
 && +  \frac{1}{8} M_{ma} E_{na} A_{\mu \nu }{}^m (A_{\mu \nu } + \gamma
_5 \tilde{A}_{\mu \nu })^n + h.c. - \nonumber \\
 && - \frac{1}{4} \bar{\Psi }_\mu  (A^{\mu \nu } - \gamma _5
\tilde{A}^{\mu \nu })^m
\bar{E}_{ma} \bar{\tau }^a \Psi _\nu  - \frac{1}{2} \bar{\lambda }_\alpha
\gamma ^\mu  \gamma _\nu  {\cal D}_\nu
z^\alpha  \Psi _\mu  + \nonumber \\
 && + \frac{i}{8} \bar{\lambda }^\alpha  \gamma ^\mu  \varepsilon _{\alpha
\beta } z^\beta  (\sigma A)^m E_{ma} \bar{\tau }^a \Psi _\mu ,
\end{eqnarray}
where $M_{ma}$ is also the function of $z^\alpha $. This Lagrangian will
be invariant under the supertransformations provided:
\begin{equation}
 \varepsilon ^{\alpha \beta } K_\alpha {}^{ma} K_\beta {}^{na} = 0, \qquad
M_{ma} \bar{K}_{na} +
\bar{M}_{ma} K_{na} = - 2 \delta _{mn}.
\end{equation}
If one chooses $M_{ma} = z^\alpha  M_{\alpha ,ma}$, where $M_{\alpha ,ma}$
is a constant matrix, satisfying $M_{\alpha ,ma} = - \varepsilon _{\alpha
\beta } \bar{M}^\beta {}_{ma}$, then the second relation gives:
\begin{equation}
 \bar{M}^\alpha {}_{ma} K_{\alpha ,na} = \delta _{mn}.
\end{equation}
Note, that this relation determines $M_{ma}$ only up to the change
$M_{ma} \to  M_{ma} + \omega  \gamma _5 K_{ma}$, but the corresponding
Lagrangians differ by the total divergency.

   Now we are ready to consider the most general interaction of $N=4$
supergravity with the arbitrary number of vector supermultiplets, the
scalar field geometry being $SU(1,1)/U(1)\otimes O(6,m)/O(6)\otimes O(m)$.
For this purpose we shall use the same set of fields as before, but
instead of six vector fields $A_\mu {}^a$ we introduce now $(6+m)$ vector
supermultiplets $(A_\mu {}^A, \Omega _i{}^A, \Phi _a{}^A)$, $A=1,2,...m+6$.
The scalar and spinor fields satisfy the following constraints:
\begin{equation}
 \Phi _a{}^A \Phi _b{}^A = - \delta _{ab}, \qquad
 \Phi _a{}^A \Omega _{iA} = 0
\end{equation}
which correspond to the required geometry and lead to the local $O(6)$
invariance, as usual. Taking into account all the scalar and spinor
field constraints, we choose the following ansatz for the
supertransformations, generalizing our results for pure $N=4$
supergravity:
\begin{eqnarray}
 \delta \Psi _\mu  &=& 2 {\cal D}_\mu  \eta  - \frac{i}{4} (\sigma A)^M
E_{MA} \Phi _a{}^A
(g^{-1})^{ab} \tau ^b \gamma _\mu  \eta ,  \nonumber \\
 \delta A_\mu {}^M &=& (\bar{\Psi }_\mu  z^\alpha  K_\alpha {}^{MA} \Phi
_a{}^A \bar{\tau }^a \eta ) + i
(\bar{\Omega }_{iA} \gamma _\mu  z^\alpha  K_\alpha {}^{MA} \eta ) - i
(\bar{\lambda }^\alpha  \gamma _\mu  K_\alpha {}^{MA}
\Phi _a{}^A \bar{\tau }^a \eta ), \nonumber \\
 \delta \Omega ^A &=& - \frac{1}{2} [ E_{MA} (\sigma A)^M - g_{AB} \Phi
_a{}^B (g^{-1})^{ab}
\Phi _b{}^C E_{NC} (\sigma A)^N ] \eta  - i \gamma ^\mu  {\cal D}_\mu
\Phi _a{}^A \bar{\tau }^a \eta , \nonumber \\
 \delta \lambda _\alpha  &=& - \frac{1}{4} \varepsilon _{\alpha \beta }
z^\beta  (\sigma A)^M E_{MA} \Phi _a{}^A (g^{-1})^{ab}
\bar{\tau }^b \eta  - i \gamma ^\mu  {\cal D}_\mu  z^\alpha  \eta ,  \\
 \delta \Phi _a{}^A &=& (\bar{\Omega }^A \bar{\tau }^a \eta ), \qquad
\delta \bar{z}_\alpha  = 2 (\bar{\lambda }_\alpha  \eta ), \nonumber
\end{eqnarray}
where $K_\alpha {}^{MA}$ are constant matrices, $E_{MA}$ and $G_{AB}$ are
functions of $z^\alpha $ to be determined, while $g_{ab} = \Phi _a{}^A
g_{AB} \Phi _b{}^B$. The requirement of the closure of the superalgebra
on the bosonic fields leads to the following relations on them:
\begin{equation}
 \bar{K}^{MA} E_{NA} = \delta ^M{}_N, \qquad K^{MA} = g^{AB} \bar{K}^{MB},
\qquad K_\alpha {}^{MA} = \varepsilon _{\alpha \beta } \bar{K}^{\beta ,MA}
\end{equation}
where $K^{MA} = z^\alpha  K_\alpha {}^{MA}$. It is not hard to construct
the fermionic part of the corresponding Lagrangian:
\begin{eqnarray}
 L_F &=& - \frac{1}{2} \bar{\Omega }^A \gamma ^\mu  \gamma ^\nu  {\cal
D}_\nu  \Phi _a{}^A \bar{\tau }^a
\Psi _\mu  - \frac{1}{2} \bar{\lambda }_\alpha  \gamma ^\mu  \gamma ^\nu
{\cal D}_\nu  z^\alpha  \Psi _\mu  - \nonumber \\
 && - \frac{1}{4} \bar{\Psi }_\mu  (A_{\mu \nu } - \gamma _5
\tilde{A}_{\mu \nu })^M
\bar{E}_{MA} \Phi _a{}^A (\bar{g}^{-1})^{ab} \bar{\tau }^b \Psi _\nu  +
\nonumber \\
 && + \frac{i}{4} \bar{\Omega }^A \gamma ^\mu  [ E_{MA} (\sigma A)^M -
g_{AB} \Phi _a{}^B
(g^{-1})^{ab} \Phi _b{}^C E_{NC} (\sigma A)^N ] \Psi _\mu  + \nonumber \\
 && + \frac{i}{8} \bar{\lambda }_\alpha  \varepsilon ^{\alpha \beta }
\bar{z}_\beta  \gamma _\mu  (\sigma A)^M E_{MA} \Phi _a{}^A
(g^{-1})^{ab} \bar{\tau }^b \Psi _\mu  + \nonumber \\
 && + \frac{1}{4} \bar{\Omega }^A [ E_{MA} - g_{AB} \Phi _b{}^B
(g^{-1})^{bc}
\Phi _c{}^C E_{MC} ] (\sigma A)^M \bar{z}_\alpha  \varepsilon ^{\alpha
\beta } \lambda _\beta  - \nonumber \\
 && - \frac{1}{8} \bar{\Omega }^A (\sigma A)^M E_{MB} \Phi _a{}^B
(g^{-1})^{ab} \bar{\tau }^b \Omega ^A.
\end{eqnarray}
The invariance of the whole Lagrangian can be achieved with the
following form of vector fields kinetic terms:
\begin{eqnarray}
 L_V &=& - \frac{1}{4} A_{\mu \nu }{}^M A_{\mu \nu }{}^N E_{MA}
\bar{E}_{NA} -
\frac{1}{4} \gamma _5 A_{\mu \nu }{}^M \tilde{A}_{\mu \nu }{}^N [
\bar{E}_{MA}
\bar{F}_{NA} - E_{MA} F_{NA} ] + \nonumber \\
 && + \frac{1}{4} A_{\mu \nu }{}^M (A_{\mu \nu }{}^N + \gamma _5
\tilde{A}_{\mu \nu }{}^N)
[E_{MA} \Phi _a{}^A (g^{-1})^{ab} \Phi _b{}^B E_{NB} ] + h.c.,
\end{eqnarray}
where $F_{MA}$ --- one more function of $z^\alpha $, provided:
\begin{equation}
 g_{AB} = E_{NA} K^N{}_B \qquad E_{MA} \bar{E}_{NA} = E{_NA}
\bar{E}_{MA} \qquad E_{MA} F_{NA} = E_{NA} F_{MA}
\end{equation}

   Thus we have managed to construct the most general (as far as we
know) model for the $N=4$--matter interaction, all previously
known results \cite{Ber85,Roo85,Roo85a,Roo85b} being partial cases of
our general formulas. Moreover, just as in the previous cases, one can
obtain as one more interesting partial case our model \cite{Tso94a},
corresponding to the interaction of arbitrary number of vector
multiplets with the hidden sector \cite{Zin87}, admitting spontaneous
supersymmetry breaking without a cosmological term.

   Note, that another interesting result can be easily obtained from
the formulas given above. Indeed, if one puts $i=1,2$, $a=1,2$,
$A+1,2,...m+2$ and uses $\tau ^a = (\varepsilon ^{ij}, \gamma _5
\varepsilon ^{ij})$ then the same
formulas given one the most general dual version for the $N=2$
supergravity interacting with vector multiplets with scalar field
geometry $SU(1,1)/U(1)\otimes O(2,m)/O(2)\otimes O(m)$!

\section{Conclusion}

   Thus, we have seen that for all extended $N=2,3,4$ supergravities
there exist dual versions for the Lagrangian of supergravity ---
matter interaction having the same scalar field geometry but different
vector fields couplings. This fact appears to be tightly connected
with the problem of spontaneous supersymmetry breaking in such
theories. The reason is that the choice of dual version determines the
global symmetry of the Lagrangian, which in turn implies different
possible gaugings and leads to the models with or without spontaneous
supersymmetry breaking, cosmological term and so on. One more
interesting question arises in the superstring context. Namely, if one
has a four dimensional superstring having the extended $N>1$
supersymmetry it is not enough to determine the geometry of the scalar
fields to fix the effective low-energy Lagrangian. As we have seen,
one has also to know the dual version it corresponds to.

\vspace{0.3in}
{\large \bf Acknowledgments}
\vspace{0.2in}

   Work is supported by the International Science Foundation grant RMP000
and
by the Russian Foundation for Fundamental Research grant 94-02-03552.

\end{document}